\documentclass[aps,prl,showpacs,twocolumn,superscriptaddress]{revtex4-1}
\usepackage[utf8]{inputenc}
\usepackage[american,british]{babel}
\usepackage[T1]{fontenc}
\usepackage[pdftex]{graphicx}  
\usepackage{graphicx, xcolor}
\usepackage{dcolumn}
\usepackage{bm}
\usepackage{amsmath,amsthm,amssymb}
\usepackage{color}
\usepackage{verbatim}
\usepackage{algorithm}
\usepackage[noend]{algpseudocode}

\definecolor{darkGreen}{RGB}{0,110,0}
\definecolor{darkBlue}{RGB}{0,0,130}
\usepackage[colorlinks,citecolor=darkGreen,linkcolor=darkBlue,%
urlcolor=blue,hyperindex]{hyperref}

\newcommand{\bra}[1]{\left\langle #1 \right|}
\newcommand{\ket}[1]{\left| #1 \right\rangle}

\begin{document}
\title{Entanglement guided search for parent Hamiltonians}

\author{X. Turkeshi}
\affiliation{The Abdus Salam International Centre for Theoretical Physics, strada Costiera 11, 34151 Trieste, Italy}
\affiliation{SISSA, via Bonomea 265, 34136 Trieste, Italy}
\affiliation{INFN, sezione di Trieste, 34136 Trieste, Italy}
\author{T. Mendes-Santos}
\affiliation{The Abdus Salam International Centre for Theoretical Physics, strada Costiera 11, 34151 Trieste, Italy}
\author{G. Giudici}
\affiliation{The Abdus Salam International Centre for Theoretical Physics, strada Costiera 11, 34151 Trieste, Italy}
\affiliation{SISSA, via Bonomea 265, 34136 Trieste, Italy}
\affiliation{INFN, sezione di Trieste, 34136 Trieste, Italy}
\author{M. Dalmonte}
\affiliation{The Abdus Salam International Centre for Theoretical Physics, strada Costiera 11, 34151 Trieste, Italy}
\affiliation{SISSA, via Bonomea 265, 34136 Trieste, Italy}

\date{\today}

\begin{abstract}
We introduce a method for the search of parent Hamiltonians of input wave-functions based on the structure of their reduced density matrix. The two key elements of our recipe are an ansatz on the relation between reduced density matrix and parent Hamiltonian that is exact at the field theory level, and a minimization procedure on the space of relative entropies, which is particularly convenient due to its convexity. 
As examples, we show how our method correctly reconstructs the parent Hamiltonian correspondent to several non-trivial ground state wave functions, including conformal and symmetry-protected-topological phases, and quantum critical points of two-dimensional antiferromagnets described by strongly coupled field theories. Our results show the entanglement structure of ground state wave-functions considerably simplifies the search for parent Hamiltonians.
\end{abstract} 

\maketitle

\paragraph{Introduction. -- } Variational wave functions have played a pivotal role in boosting the understanding of strongly correlated systems~\cite{wenbook,greiterBook,fradkinbook,Laughlin,PhysRevB.59.8084,Moore1991362}. The success of ansatz wave functions has naturally motivated the search for the corresponding parent Hamiltonians, with considerable success in several contexts, including the study of topological matter~\cite{Haldane:1983aa,Kapit:2010aa,Grover_2013}, one-dimensional systems~\cite{Haldane:1992aa,Affleck1987}, and tensor networks~\cite{schollwock2011density,SCHUCH20102153}. Recent experimental progresses in quantum engineering of synthetic systems~\cite{Bloch2012,blatt2012quantum,Buluta2011,Georgescu2014,Cirac2012} have opened an additional perspective in the search for parent Hamiltonians: thanks to the high degree of interactions tunability, these experiments provide a clean route toward the realization of tailored quantum dynamics. This has stimulated a renewed theoretical interest as of late. In Refs.~\cite{Swingle_2014,Garrison_2018,1712.01850,1802.01590,1807.04564}, a series of approaches has been proposed for determining, given an initial quantum state $|\Psi\rangle$, a Hamiltonian operator $H$ which has $|\Psi\rangle$ has an eigenvector, very remarkably, even utilizing limited information such as low-order correlation functions~\cite{1712.01850,1802.01590,1807.04564}. However, it remains unclear if a generic procedure exists to determine an (approximate) Hamiltonian operator that has $|\Psi\rangle$ as its {\it ground state}~\cite{vishwanath2018}.

\begin{figure}
\includegraphics[width=0.85\columnwidth]{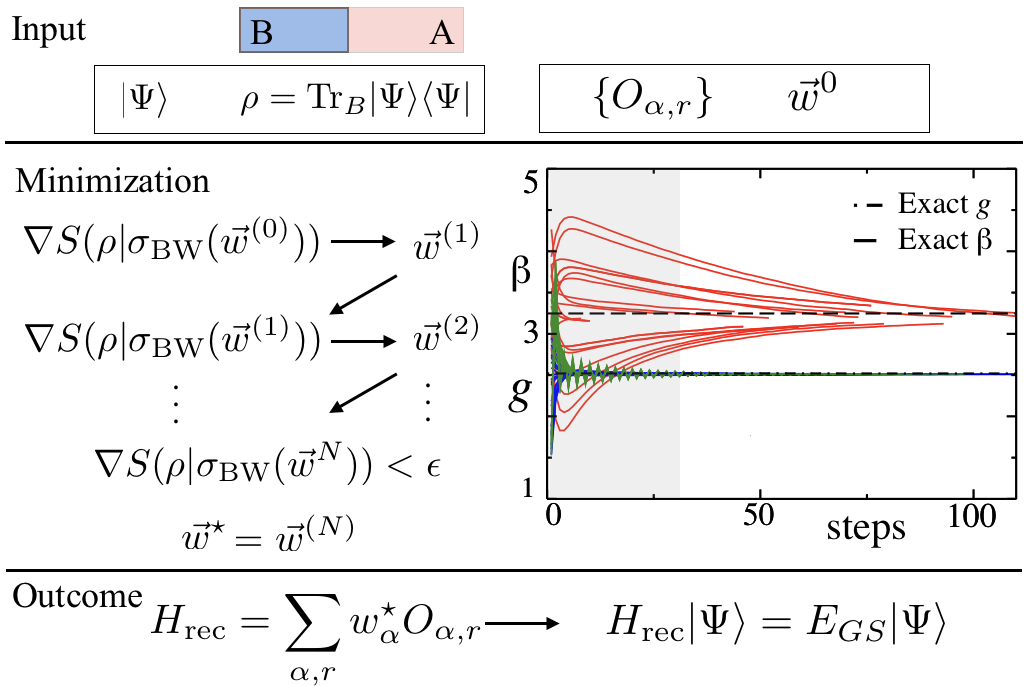}
\centering
\caption{ Schematics of the parent Hamiltonian search. The starting point is the wave-function of interest $|\Psi\rangle$ (with a half-partition reduced density matrix $\rho$), a set of local operators $\{O_{\alpha,r}\}$, and an initial guess for their coefficients $w_\alpha^{0}$. The relative entropy $S$ between $\rho$ and an ansatz Bisognano-Wichmann reduced density matrix $\sigma (\{O_{\alpha,r}, w_\alpha^{n}\})$ is evaluated at each step of the minimization procedure. The procedure is then stopped once the desired accuracy ($\epsilon$) is reached: the final outcome are the couplings $w_\alpha^*$ of the parent Hamiltonian, and the entanglement inverse temperature $\beta$. 
The inset shows a sample of our results for the bilayer Heisenberg model at the critical point (see text). Different lines corresponds to minimization from different initial sets $\vec{w}_0$. Convergence within 0.1\% of the correct value of $g$ (interlayer coupling; blue and green) and $\beta$ (red lines) is typically reached after 30 (grey area) and $\simeq$200 steps, respectively.
}
\label{fig:scheme}
\end{figure} 

In this work, we introduce a method to determine parent Hamiltonians guided by field theoretical constraints between the parent Hamiltonian itself, and the entanglement structure of $|\Psi\rangle$. The first key element of our strategy is to connect the reduced density matrix $\rho$ of the half-partition of $|\Psi\rangle$ to an ansatz given by the Bisognano-Wichmann theorem~\cite{bisognano1975duality,bisognano1976duality,haag2012local,Witten:2018aa} adapted to the lattice~\cite{Dalmonte:2017aa}. As we detail below and schematically illustrate in Fig.~\ref{fig:scheme}, this provides an immediate connection between the input vector $|\Psi\rangle$ and its translational invariant parent Hamiltonian $H$ through $\rho$. The second key element is to combine this ansatz to a minimization procedure based on relative entropy~\cite{Vedral:2002aa}: this guarantees that the target space of potential parent Hamiltonians is convex (even if no statement can be made {\it a priori} on the convergence rate to the right solution), and its volume is system size independent. The entire procedure can be carried out using different theoretical methods that rely on evaluating expectation values at finite (entanglement) temperature, including finite-temperature density-matrix-renormalization-group (TDMRG)~\cite{White1992,white2005} and quantum Monte Carlo simulations~\cite{Kaul_2013}, as we explicitly demonstrate.

Our method is applicable to a broad range of wave functions which can be thought of as ground states of lattice Hamiltonians described at low energies by quantum field theories with emergent Lorentz invariance - thus encompassing physical phenomena as diverse as quantum criticality with $z=1$, topological matter, and quantum antiferromagnets, to name a few. We illustrate the effectiveness of the procedure and discuss its scaling to the right solution by considering three examples, which encompass qualitatively different phenomena: conformal phases in the one-dimensional (1D) XXZ chain~\cite{fradkinbook}; the Haldane chain as an epitome of a symmetry-protected topological phase~\cite{haldane1983nonlinear}; and eventually, an example of strongly coupled quantum critical point in the two-dimensional (2D) bilayer Heisenberg antiferromagnet~\cite{Chubukov:1994aa,sandvik1994,Wang:2006aa}. A sample of our results for the latter model is depicted in Fig.~\ref{fig:scheme}: we plot the convergence of the microscopic parameters - the ratio between inter- and intra- layer couplings $g$, and the inverse entanglement temperature $\beta$ - as a function of the number of steps in the minimization. Convergence to within $10^{-3}$ (see below) of the exact results for $g$ is typically achieved in around 10 steps.

\paragraph{Entanglement Hamiltonian structure. -- } The parent Hamiltonian search we propose is guided by a set of field theory results, which go under the name of Bisognano-Wichmann (BW) theorem~\cite{bisognano1975duality,bisognano1976duality,haag2012local,Witten:2018aa}, which we briefly review. While it is well known that for local Hamiltonians the ground state contains a (potentially complete) information about the Hamiltonian spectrum, this theorem allows to quantitatively establish this correspondence at the field theoretic level.

Given a pure state $\ket{\Psi}$ and a bipartition $A\cup B$, one can re-express the reduced density matrix as:
\begin{equation}
\label{BWft}
	\rho=\text{Tr}_{B}\ket{\Psi}\bra{\Psi}=\frac{e^{- H_A}}{Z_A},\quad Z_A=\text{Tr}_{A}e^{- H_A}.
\end{equation}
The operator $H_A$ is called entanglement (or modular) Hamiltonian~\cite{laflorencie2016,Witten:2018aa} and, its spectrum is known as entanglement spectrum~\cite{Regnault:2015aa}. The BW theorem states that if $\ket{\Psi}$ is the vacuum state of a relativistic quantum field theory defined in the thermodynamic limit by an Hamiltonian density $h(x)$, and the bipartition is over half-space (e.g. in $D+1$ dimensions $A=\{\vec{x}|x_1>0\}$), then:
\begin{equation}\label{BWH}
	H_A =\beta\int_{x_1>0} x_1 h(\vec{x})d^Dx
\end{equation}
The parameter $\beta$ is a prefactor related to the sound velocity of the theory; it is referred to as the inverse entanglement temperature.
Recently, based both on exact analytical results and growing numerical evidence~\cite{Itoyama:1987aa,peschel2009,Kim_2016,1806.08060,Toldin:2018aa,Dalmonte:2017aa,eisler2018,Kosior:2018aa,Giudici:2018aa}, it has been argued that these results can be applicable to obtain very accurate (if not exact) approximations of entanglement Hamiltonians of lattice models, as long as their low-energy physics is effectively described by a relativistic quantum field theory. While not generic, this common structure encompasses a plethora of phenomena in the field of strong correlations, including quantum critical points and phases with emergent relativistic description. As a case sample, in 1D, this reads	 $H_A= \beta\sum_{n>0} n \ h_n$, which is the discretized version of the field theory result, where $h_n$ are the local (i.e., defined on sites and on bonds) terms of the lattice Hamiltonian. The result is trivially extended to D>1.

\paragraph{Parent Hamiltonian search algorithm. -- } Equipped with the direct relation between $|\Psi\rangle$ and the system Hamiltonian provided by the discretized version of Eq.~\eqref{BWH}~\cite{Dalmonte:2017aa}, we formulate now our search algorithm. 

{\it 1) Input:} Given a lattice input state $\ket{\Psi}$ and a local basis of hermitian operators $\{O_{\alpha,r}\}$ labelled by a lattice index $r$ and an internal index $\alpha$, our goal is to find the coefficients $\vec{w}^\star$ of the linear combination:
\begin{equation}
\label{eq.1}
	H(\vec{w}) = \sum_{\alpha} w_{\alpha,r}O_{\alpha,r},
\end{equation}
such that the input state is its ground state. We call this local operator the reconstructed Hamiltonian ${H_\textup{rec}=H(\vec{w}^\star)}$. Here, we focus on translationally invariant Hamiltonians, and thus set ${w_{\alpha,r}=w_{\alpha}}$.

{\it 2) Minimization:} In order to construct $H_{\text{rec}}$, we propose an optimization procedure based on minimizing the relative entropy \cite{PhysRevX.8.021050,1803.11278} utilizing as trial reduced density matrix the BW one. The relative entropy between two density matrices $\rho$ and $\sigma$ is defined as~\cite{Vedral:2002aa,Blanco2013}:
\begin{equation}
	S(\rho|\sigma) = \text{Tr}(\rho\log\rho) - \text{Tr}(\rho\log\sigma).
\end{equation}
For our purpose here, this function has two key features: it is non-negative, i.e. $S(\rho|\sigma)\ge 0$, with the equality holding if $\rho=\sigma$; and it is joint-convex. 
This latter property ensures the uniqueness of a global minimum for the function $S(\rho|\sigma)$ with fixed $\rho$.

In the context of our problem, the left argument $\rho$ encodes the input data, that is, $\rho= \text{Tr}_{B}\ket{\Psi}\bra{\Psi}$.
The right argument $\sigma$ is the reduced density matrix of the GS of $H(\vec{w})$, which returns the parent Hamiltonian at the end of the procedure. We thus express $\sigma$ by using the Bisognano-Wichmann density matrix in eq.~\eqref{BWft}:
\begin{equation}
	\sigma_\textup{BW}(\vec{w}) = \frac{e^{-H_\textup{BW}(\vec{w})}}{Z_A(\vec{w})};\quad Z_A(\vec{w})= \text{Tr}_Ae^{-H_\textup{BW}(\vec{w})},
\end{equation}
with $H_A$ of the form:
\begin{equation}
	H_\textup{BW}(\vec{w})=  \sum_\alpha\sum_{r>0} w_\alpha r O_{\alpha,r}= \sum_\alpha  w_\alpha h_\alpha.
\end{equation}
{\it 3) Outcome: }Under the assumptions above, the parent Hamiltonian is given by the set of parameters that uniquely minimize the relative entropy; that is, the coefficients:
\begin{equation}
	\vec{w}^\star = \arg\min_{\vec{w}} S(\rho|\sigma_\textup{BW}(\vec{w})).
\end{equation}
in combination with Eq.~\eqref{eq.1}, determine the parent Hamiltonian. The minimization procedure of $S(\rho|\sigma)$ can be carried out in several ways: below, 
we utilize adaptive gradient descent methods (see Ref.~\cite{supmat} for details).
Given an initial configuration, we carry out the minimization of $S(\rho|\sigma)$ by evaluating the \textit{error}
\begin{equation}
  \epsilon = || \eta \nabla S(\rho|\sigma_\textup{BW}(\vec{w}))||,
  \label{error}
\end{equation}
where $\eta$ is a control parameter (see Ref.~\cite{supmat} for details),
until convergence to the given accuracy in the coupling parameters of $H_{BW}$ is reached~\footnote{For unknown problems, 
one shall perform convergence in $S(\rho|\sigma)$, which behaves similarly to the couplings.}. For the sake of convenience, we consider here a $10^{-3}$ error threshold, that already returns a very accurate parent 
Hamiltonian, while in Ref.~\cite{supmat} we show results down to $10^{-9}$. This passage does not require access to the wave function, but is rather carried out evaluating the expectation value 
of local correlators at finite (entanglement) temperature: as such, it is immediately amenable to a series of methods, including Monte Carlo - as we show below. Since the minimization space has constant 
dimension with system size, and since it is convex, we expect a mild - if any - scaling with system size of the time to solution.

\begin{figure}[t]
\centering
\includegraphics[width=0.95\columnwidth]{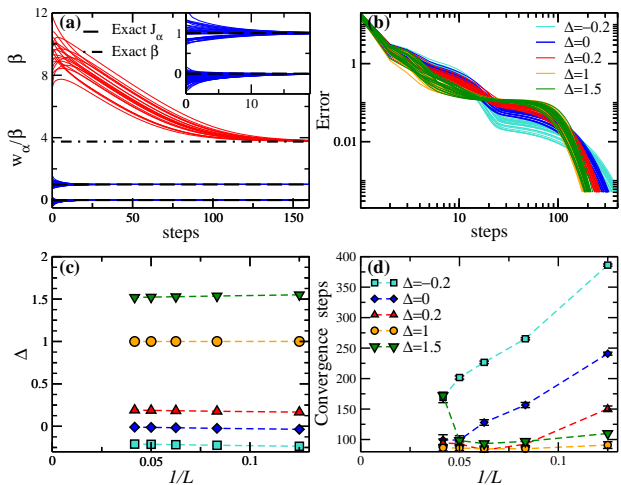}
\caption{Parent Hamiltonian search for the XXZ chain. {({a})} Ratios $J_\alpha=w_\alpha/\beta$ (blue lines) and $\beta$ (red lines) as functions of the minimization steps. We compare with the exact coupling and entaglement temperature. Here, $L=12$ and $\Delta=1$. We consider the full basis of operators $\{w_\alpha\}$ (see text).
{({b})} Error, Eq.~\eqref{error}, as function of the minimization steps; the value $\Delta$ label the different input state used, ground state of the respective XXZ Hamiltonian. 
{({c})} Average $\Delta$ over initial realizations of the coupling as a function of system size. Here and in panel (d), the considered parameters are restricted to $\beta$ and ${\Delta=w_{zz}/\beta}$.
{({d})} Average convergence steps over initial realizations of the coupling as a function of system size. The convergence rate typically improves with system size.
}
\label{fig:xxz.5}
\end{figure} 

Before proceeding, it is worth pointing out that the method is not immediately suited to simple wave functions, where correlations vary at the lattice spacing level (like a product state). This is due to the field theoretical input we employ, which might fail in these regimes. Another potential limitation is that it is not possible to capture parent Hamiltonians with quadratic spectra, such as ferromagnets. Failure is in principle straightforward to diagnose - the relative entropy minimum will attain a large value, indicating the result is not correct. In the following, we purposely benchmark our strategy focusing purposely on non-trivial wave functions which {\it lack} simple tensor network representations.

\paragraph{Parent Hamiltonian of conformal phases. -- } The class of wave functions we consider are $c=1$ CFTs (Luttinger liquids); specifically, we consider the ground states of the XXZ s-1/2 chain, defined as:
\begin{equation} \label{xxz}
H=\sum_{\langle i,j\rangle} \left(S^x_i S^x_{j} +S^y_i S^y_{j} + \Delta S^z_{i} S^z_{j}  \right),
\end{equation}
where $\langle \cdot\rangle$ is the restriction to nearest neighbor terms, and $S^\alpha_i$ are spin-1/2 operators at the site $i$. The model hosts a gapless phase for $-1<\Delta\leq 1$, described at low-energies by a $c=1$ CFT (Luttinger liquid); in addition, it displays a ferromagnetic (antiferromagnetic) phase for $\Delta\leq-1$($\Delta>1$). Both the gapless and the antiferromagnetic phase shall be captured by our approach.

\begin{figure}[t]
\includegraphics[width=0.75\columnwidth]{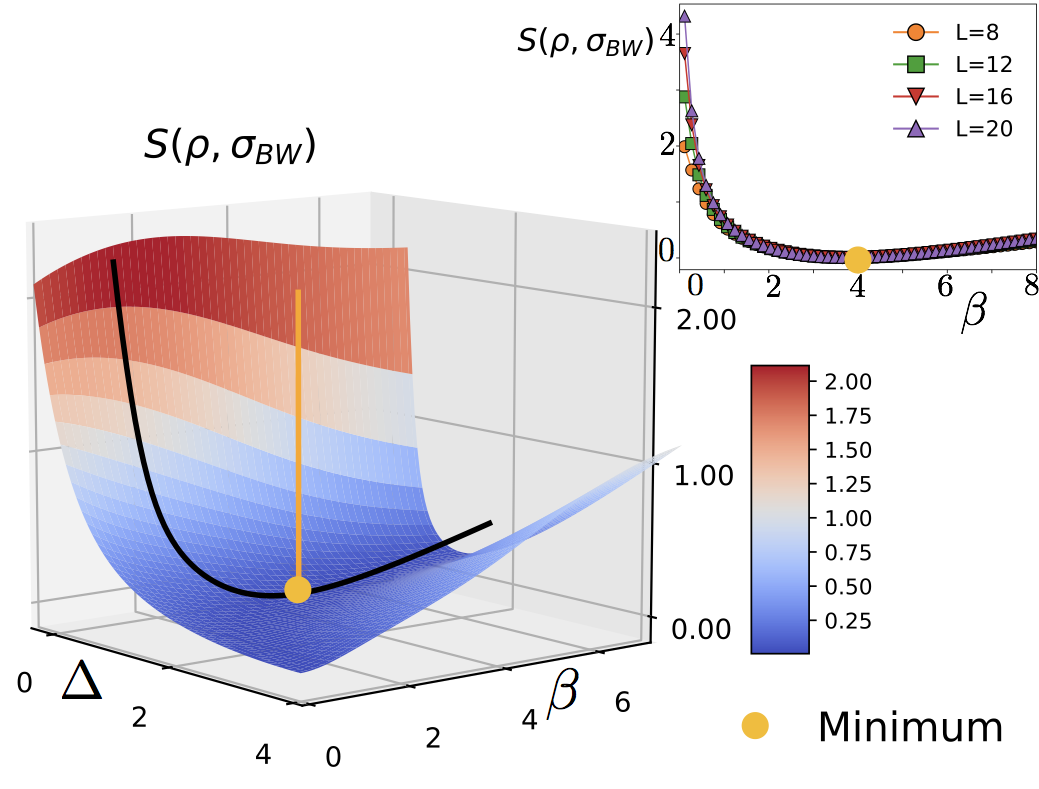}
\centering
\caption{Landscape of the relative entropy landscape between $\rho$ (obtained from the ground state at $\Delta=1$) and $\sigma_{BW}(\Delta, \beta)$ for $L=12$: the minimum is at $(1., 4.)$. The insets shows a planar cut ($\Delta=1$) for different system sizes.
}
\label{fig:relentr}
\end{figure} 

To test our method, we choose as ansatz wavefunctions the ground state of Eq.~\eqref{xxz} for various values of $\Delta$, and as the basis of operators $\{O_{\alpha,r}\}=\{S^{a}_{r} S^{b}_{r+1},S^{a}_{r}\}$ with $a,b=x,y,z$ and $r$ the lattice-site label. We define the convergence parameters as ${w_\alpha=\beta J_\alpha}$, and consider open boundary conditions (as in the examples below).

Fig.~\ref{fig:xxz.5} shows the outcomes of our algorithm using exact diagonalization (ED) up to a total system size $L=12$ (panel (a,b)) and TDMRG~\cite{white2005} up to $L=24$ (panel ({c,d})). In panel (a), we plot the $w_\alpha$ as a function of the steps for different initial guesses $\vec{w}_0$: the symmetries of the systems are rapidly identified (unwanted terms vanishing), and the couplings of the parent Hamiltonian converge to the correct ratios after few steps; the entanglement temperature converges slower. The relative entropy indicating vicinity to the exact solution (Fig.~\ref{fig:xxz.5}b) displays few plateaus, and eventually converges (exponentially fast) to the correct solution.

In Fig.~\ref{fig:xxz.5}c, we plot the converge of $\Delta$ to the correct result as a function of system size. We remark that scaling with system size is not trivial for critical systems, due to the structure of reduced density matrices~\cite{Wong2013,Lashkari:2014aa}(see Ref.~\cite{supmat} for an extended discussion). Finally, in Fig.~\ref{fig:xxz.5}d we plot the number of convergence steps needed to reach $\epsilon=10^{-3}$ threshold using only two free parameters for simplicity: remarkably, the procedure becomes simpler when increasing system size. The case $\Delta=1.5$ shows an abrupt increase at $L=24$: this is an artefact of the minimization and it is easily removed (see Ref.~\cite{supmat} for details). These results are fully consistent with the relative entropy landscape depicted in Fig.~\ref{fig:relentr}

\begin{figure}[t]
\includegraphics[width=0.99\columnwidth]{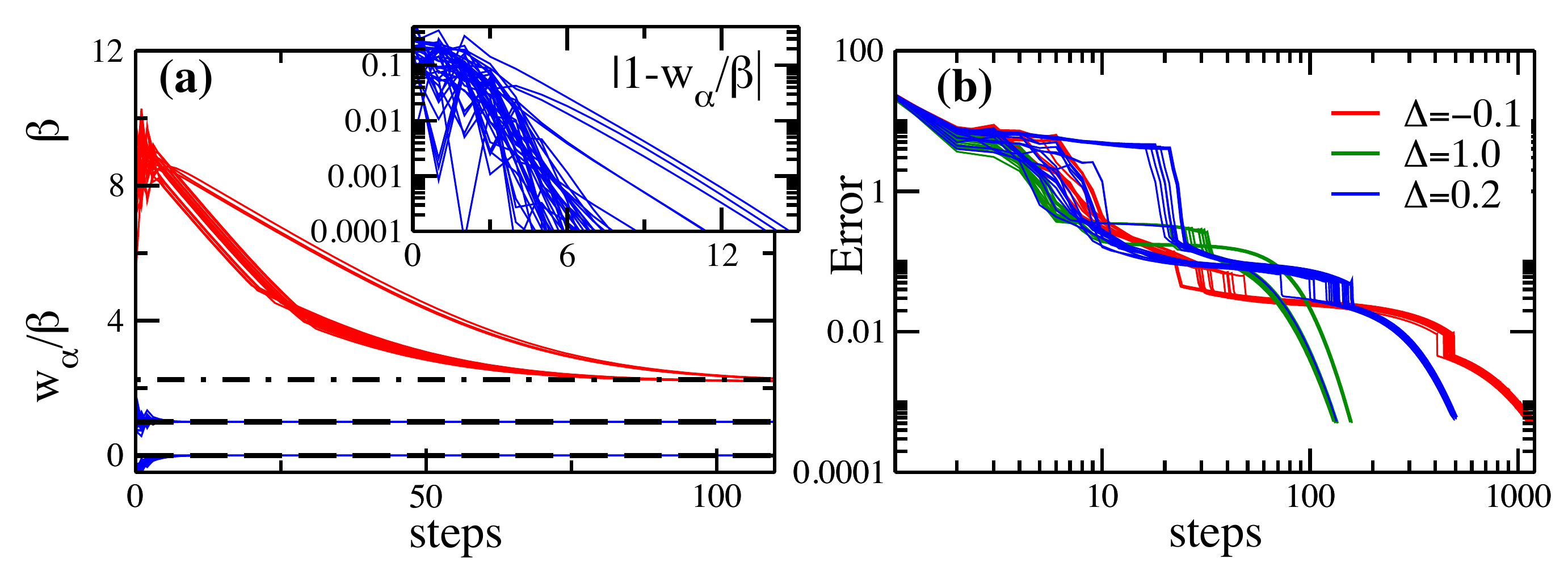}
\centering
\caption{Parent Hamiltonian search for the Haldane chain with $L=8$. (a) $J_\alpha=w_\alpha/\beta$ and $\beta$ as functions of the minimization steps; the initial state is at $\Delta=1$. The inset shows a magnification up to 14 steps of $|1-w_\alpha/\beta|$: convergence to the correct solution at $10^{-4}$ level  is typically achieved after 6 steps. (b) Error as function of minimization steps for different values of $\Delta$.}
\label{fig:xxz1}
\end{figure}

\paragraph{Parent Hamiltonian of a symmetry-protected topological phase. -- } As a model with non-trivial topological phase, we discuss here the Haldane chain~\cite{haldane1983nonlinear,fradkinbook}, described by the Hamiltonian Eq.~\eqref{xxz} with spin-1 operators. For $0<\Delta\lesssim1.2$, the model supports a Haldane phase~\cite{haldane1983nonlinear,Chen:aa}. We used ED with adaptive gradient descent to determine the parent Hamiltonian for different values of $\Delta$. In this case, we have chosen a subset of the full basis of local hermitian operators up to two body terms $\{O_{\alpha,r}\}=\{S^a_{r} S^b_{r+1},S^a_{r}\}$ (i.e., we do not include spin-1 local operators as $(S^{\alpha}_{r})^2$). The results of the minimization procedure are illustrated in Fig.~\ref{fig:xxz1}: in full analogy with the $s$-1/2 case, the couplings quickly converge to the correct results, while $\beta$ convergence is slower. In all instances we studied, the relative entropy converged faster to 0 in the gapped, topological regime (Fig.~\ref{fig:xxz1}b),

\begin{figure}
\includegraphics[width=0.95\columnwidth]{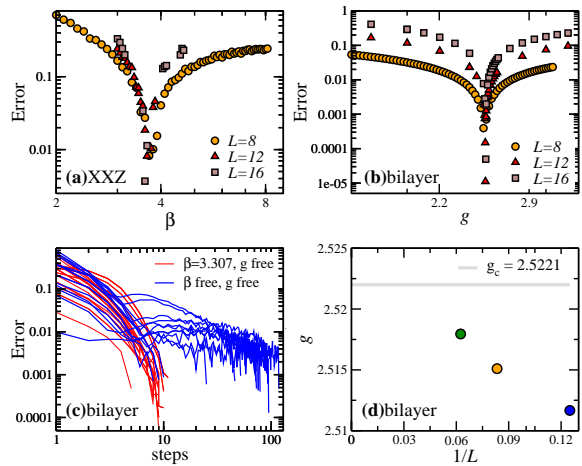}
\centering
\caption{(a-b) Error between the exact and the BW reduced density matrix, Eq.~\eqref{error}, as a function of temperature for the 2D Heisenberg model (a), 
and as a function of $g$ for the bilayer Heisenberg model at $\beta=3.3$ (b).  (c)Error as a function of the minimization step for $L=8$; the error decreases quickly, without any plateau as in the 1D case. (d) Converged inter-to-intra layer coupling versus linear system size.}
\label{fig:2DHeis}
\end{figure}

\paragraph{Parent Hamiltonian of a two-dimensional quantum critical point. -- } In our third example, we consider 2D antiferromagnets described by Eq.~\eqref{xxz}, that we treat using a stochastic series expansion 
Monte Carlo including loop updates~\cite{sandvik1991,sandvik2002}. In order to check feasibility of the approach in 2D (whose bipartition size is $L/2\times L/2$), 
we investigated convergence to the correct entanglement temperature in the Heisenberg model (without changing the Hamiltonian parameters). 
As discussed in the supplementary material, convergence was achieved typically after few tens steps. We plot in Fig.~\ref{fig:2DHeis}a the error $\epsilon$ at fixed $\beta$: 
as expected, this correctly features a minimum at the right value $\beta=2\pi/v=3.792$~\cite{wiese1998}, weakly dependent on system size as expected~\cite{Giudici:2018aa}. 
The landscape is sharper at larger $L$, indicating faster convergence.

We then tested our approach trying to reconstruct the correct parent Hamiltonian for the bilayer Heisenberg model, 
characterized by the ratio of inter-to-intra layer coupling $g$ \cite{sandvik1994,Wang:2006aa}. 
In particular, we focused on its critical point, which separates a disordered and an antiferromagnetic phase, and is located at $g_c=2.52210(5)$ \cite{Wang:2006aa,PhysRevB.92.195145}. 
At this point, the system dynamics is described by a $\sigma$-model~\cite{Chubukov:1994aa,Wang:2006aa,Kaul_2013}. 
Before applying our procedure, we performed a preliminary check on the relative entropy manifold as a function of the coupling $g$ at fixed $\beta$, see Fig.~\ref{fig:2DHeis}b: 
the minimum of the error clearly signals the correct coupling. 

We then applied two different procedures: in the first one, we fixed the entanglement temperature, and let the coupling $g$ free (red lines in Fig.~\ref{fig:2DHeis}c), while in the second, we let both $g$ and $\beta$ vary (blue lines). In both cases, the error $\epsilon$ quickly diminishes (slower in the second case due to more parameters to be optimized). Most importantly, at the end of the minimization, the value of $g$ is extremely close to the correct one, which seems to be correctly reproduced in the thermodynamic limit (Fig.~\ref{fig:2DHeis}d). Given the complexity of the system wave function, this serves as a strong benchmark for our strategy: reconstructing the parent Hamiltonian in this case takes only few tens of steps, each one corresponding to a MC simulation of the BW entanglement Hamiltonian.

\paragraph{Conclusions and outlook. --} We proposed a method to guide the search of parent Hamiltonians utilizing insights on the entanglement structure of ground state wave functions based on the Bisognano-Wichmann theorem adapted to the lattice. We benchmarked the feasibility of our strategy utilizing several input wave functions, finding convergence to the correct solution in a number of steps that typically decreases with system size. It will be interesting to check whether connections to restricted Boltzmann machine methods~\cite{1803.11278} can be established when inputing finite temperature states, where, in certain cases, the form of the entanglement Hamiltonian can be derived~\cite{Cardy:2016aa}, and how to extend the method to gauge theories, which play a pivotal role in the understanding of spin liquids~\cite{becca2011,Grover_2013}, and whose Hilbert space structure requires a more refined approach~\cite{Buividovich:2008aa,Casini:2014aa}.

\paragraph{Acknowledgement. - }

We acknowledge useful discussions with P. Calabrese, R. Fazio, A. Scardicchio, and E. Tonni. MD is grateful to A. Vishwanath for discussions stimulating this work at the Simons Conference on Ultra Quantum Matter. 
This work is supported by the ERC under grant number 758329 (AGEnTh). 
TMS and MD acknowledge computing resources at Cineca Supercomputing Centre through the Italian SuperComputing Resource Allocation via the ISCRA grants TopoXY and QMCofEH.

\phantomsection

\bibliography{HpsiBib.bib}

\widetext
\clearpage

\onecolumngrid
\begin{center}
  \textbf{\Large Supplementary Material: \\ \medskip
\Large Entanglement guided search for parent Hamiltonians }\\[.2cm]
\end{center}

\twocolumngrid

\setcounter{equation}{0}
\setcounter{figure}{0}
\setcounter{table}{0}
\setcounter{page}{1}
\renewcommand{\theequation}{S\arabic{equation}}
\renewcommand{\thefigure}{S\arabic{figure}}


\section{1D models: XXZ and Haldane chain.}
\label{1d}
In this section, we provide additional details about the implementation of our strategy on the 1D models discussed in the main text:
\begin{equation}
\label{eq11}
	H=\sum_i \left(S^x_i S^x_{i+1} +S^y_i S^y_{i+1} + \Delta S^z_{i} S^z_{i+1}  \right),
\end{equation}
where the operators $S_{\alpha,r}$ are the usual Pauli matrices for the XXZ chain, while the spin-1 $SU(2)$ representation for the Haldane chain. Both these instances have effective low energy relativistic quantum field theory description, thus they are good tests for our algorithm, which relies on the Bisognano-Wichmann (BW) characterization of the ground state reduced density matrix. We remark that, for the spin-1/2 case, there is also a deeper connection between the BW theorem itself and the lattice realization of boost operators~\cite{Itoyama:1987aa}.
\begin{figure}[h!]
	\centering
	\includegraphics[width=0.8\columnwidth]{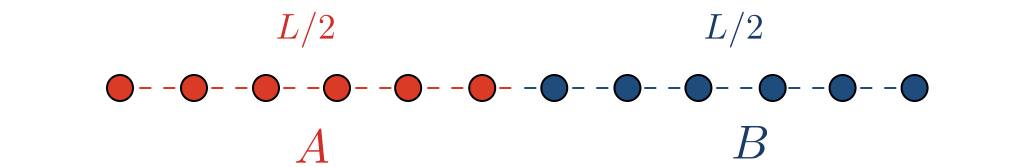}
\end{figure}
Given a half-partition of the system $A\cup B$, we compute the reduced density matrix over $A$ of the ansatz state $\Psi$ (in our examples the ground state of eq.~\eqref{eq11}) and we fit this with a model entanglement Hamiltonian (EH) of the BW type:
\begin{eqnarray}
	H_\textup{BW} &=& \sum_{\alpha,r>0} w_\alpha r O_{\alpha,r} 
\\ 	&=& \sum_{a,b=x,y,z}\sum_{r>0} r\left(w_{a b}S_{a,r}S_{b,r+1}+w_{a} S_{a,r}\right).
\end{eqnarray}

\begin{figure}[tb]
	\centering
	\includegraphics[width=\columnwidth]{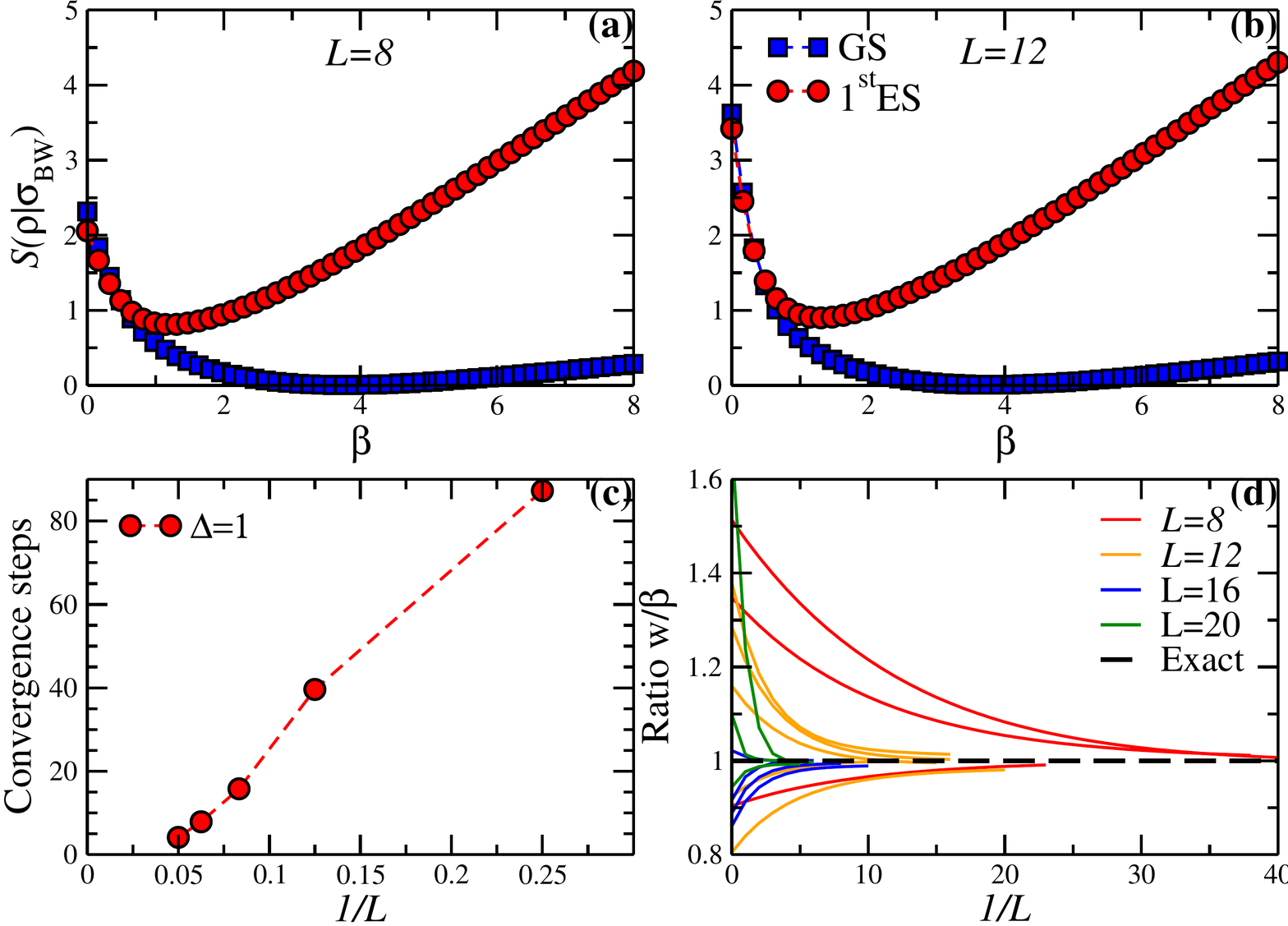}
	\caption{(a,b)Relative entropy of the ground state and of the first excited state with respect to the Bisognano-Wichmann density matrix for $L = 8,12$ and $\Delta=1$. This landscape is weakly affected by the finite system size, as discussed in the main text. (c)  Average convergence steps of the ratio $w_{xx}/w_{zz}$ as function of the system size. We focus on the symmetric situation of the XXZ model $w_{xx}=w_{yy}\neq w_{zz}\neq 0$ (other coefficient set to zero). As input state, we use the isotropic chain ground state. (d) Some representative trajectories of the coefficients at different system sizes. The setup is the same of the previous panel.}
	\label{fig:figure1}
\end{figure}

This form (Hamiltonian density with a site-dependent prefactor) uniquely characterizes the ground state properties. As a first illustration, we show in Fig.~\ref{fig:figure1} the relative entropy landscape between the BW EH correspondent to the GS and the reduced density matrix obtained from the first excited state. This shows that minimization starting from the first excited state typically does not return the original Hamiltonian at zero relative entropy (indicating that the minimization has not succeeded).

Thus, if minimization is reached  $\{w_\alpha^\star\}=\{w_{a b}^\star,w^\star_a\}$ ($a,b = x,y,z$) and the relative entropy at the minimum is close to zero (within $\epsilon$ as defined in the main text), we claim that the input state is the ground state of the reconstructed Hamiltonian:
\begin{eqnarray}
	&&H_\textup{rec} =  \sum_{a,b=x,y,z}\sum_r \left(w_{a b}^\star S_{a,r}S_{b,r+1}+w_{a}^\star S_{a,r}\right),\\
	&&H_\textup{rec} \Psi = E_{GS}\Psi.
\end{eqnarray}

Minimization is pursued by adaptive gradient descent (GD). For practical purposes, we use the notation $\partial_\alpha=\partial/\partial w_\alpha$ and:
\begin{equation}
	\langle{O}\rangle_{\textup{dat}} \equiv \text{Tr}(O \rho),\quad \langle{O}\rangle_{\text{BW},\vec{w}} \equiv \text{Tr}(O \sigma_{\textup{BW}}{(\vec{w})}).
\end{equation}
Now, given a  configuration of parameters $\vec{w}^{(n)}$, we compute the gradient of the relative entropy:
\begin{align}
\label{gd1}
	\partial_\alpha S(\rho_\textup{dat}|\sigma_\textup{BW}(\vec{w})) &= \partial_\alpha\left(\langle{H(\vec{w})}\rangle_{\textup{dat}} - \log Z_A(\vec{w}) \right)|_{\vec{w}^{(n)}} \nonumber\\
	& = \langle{\tilde{h}_\alpha}\rangle_{\textup{dat}} - \langle{\tilde{h}_\alpha}\rangle_{\text{BW},\vec{w}^{(n)}}.
\end{align}

To compute the relative entropy gradient at some value $\vec{w}^{(n)}$ we thus just need the averages of the correlation functions correspodent to the terms allowed in the EH, evaluated over the ansatz state and over the BW density matrix evaluated with $\vec{w}^{(n)}$ of the operators $O_{\alpha,r}$ (using the same notation of the main text). Thus, rewriting the previous equation in a more explicit form, we need to compute:
\begin{align}
\label{eq22}
	\partial_\alpha S(\rho|\sigma_\textup{BW}(\vec{w})) = \sum_{r>0} r \langle{O_{\alpha,r}}\rangle_{\textup{dat}} - \langle{O_{\alpha,r}}\rangle_{\text{BW},\vec{w}^{(n)}}.
\end{align}
We emphasize again that this passage does not required any knowledge of the wave function at a given step: this is a key feature in view of applying this minimization procedure in combination with Monte Carlo methods.

Here is the scheme of the algorithm:
\noindent\rule{\columnwidth}{2pt}
\begin{algorithmic}
\Procedure{Reconstruction}{$\Psi,\{O_{\alpha,r}\}$}
\State Initialization: $\vec{w}^{(0)}$, error = 10, $\eta = 4$\Comment{$\eta$ adapted along the run.}
\State Compute: $\sum_r r \langle{O_{\alpha,r}}\rangle_{\Psi}$
\While{error < threshold}
\State Compute: $\sum_r r \langle{O_{\alpha,r}}\rangle_{\vec{w}^{(n)}}$
\State Compute: $\partial_\alpha S$
\State Compute: $\vec{w}^{(n+1)} = \vec{w}^{(n)} + \eta\nabla S(\rho|\sigma_\textup{BW}(\vec{w}^{(n)}))$
\If{$||\vec{w}^{(n+1)}-\vec{w}^{(n)}||>$error $\lor ||\vec{w}^{(n+1)}-\vec{w}^{(n)}||$ is stationary}
\State $\eta = \eta/2$\Comment{If error stationary or grow, make $\eta$ smaller}
\State error = $||\vec{w}^{(n+1)}-\vec{w}^{(n)}||$\Comment{Update error.}
\EndIf
\EndWhile
\Return $H_\textup{rec}= \sum_{\alpha,r} w^{(N)}_\alpha O_{\alpha,r}$\Comment{Reconstructed Hamiltonian up to mupltiplicative constant.}
\EndProcedure
\end{algorithmic}
\noindent\rule{\columnwidth}{2pt}
Since GD requires many iterations until convergence is reached, we used exact diagonalization for small system sizes (up to $L=12$) when considering all the $12$ coefficients $ \{w_{a b},w_a\}$ in our model. While in the case in which we utilized the $U(1)$-symmetric version of the algorithm ($w_{xx}=w_{yy}\neq w_{zz}\neq 0$ and the other couplings set to zero) we were able to perform computations up to $L=24$ with exact diagonalization and finite temperature density matrix renormalization group. In the former case we can exploit magnetization conservation by computing the full eigensystem of $\sigma_{BW}$ upon restriction to all symmetry sectors. In the latter case we can still exploit $U(1)$ symmetry by preparing the purified state (system + ancilla) in an eigenstate of the total magnetization (see \cite{white2005} for details). The number of states kept during the imaginary-time-evolution, which provides the desired thermal state, was chosen to increase during the evolution, starting from $20$ up to $100$ states per block.\\ 
We considered uniform random instances of $\vec{w}^{(0)}$ over the interval $I=[2,6]$ (in order to keep computational costs cheap, the neat results are unchanged by this choice) and averaged over a hundred of initial conditions; we kept track of these by fixing the seeds of the pseudo-random number generators implemented. From now on, we discuss results for the spin-1/2 case.

\begin{figure}[h]
	\centering
	\includegraphics[width=\columnwidth]{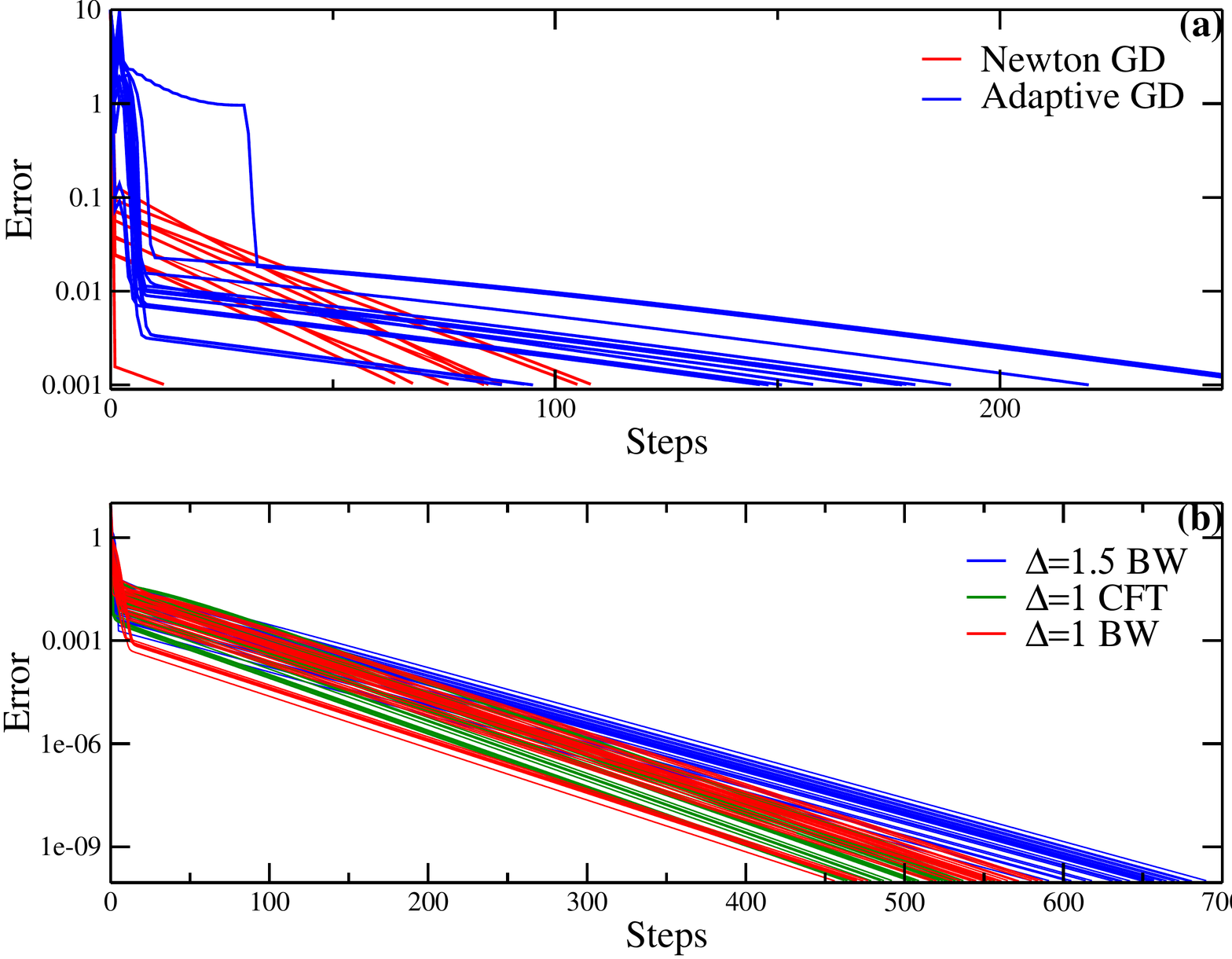}
	\caption{(a) Comparison between two different schemes for gradient descent minimization for the XXZ chain at $L=24$ and $\Delta =1.5$. (b) Error trajectories for $\Delta=1$ XXZ chain at $L=16$. For $\Delta = 1$ we used both the BW prescription and the CFT one; as it is shown, the results are similar in terms of scaling.}
	\label{fig:figure333}
\end{figure}
Our results (figure~\ref{fig:figure1}) show that the ratios of the various coefficients converge very rapidly to the correct ones, at a rate which increases with the size of the system. In particular, the ratios related to couplings which are not compatible with the input wave function symmetry (in our cases here, $U(1)$), such as $w_{xz}$, decay extremely rapidly to 0. While we have not exploited that, it is possible to devise an optimal approach where, at first, one lets the ratios correspondent to all couplings which are incompatible with the symmetry flow to zero (e.g., by running a test simulation and select those that decay faster), and then one optimizes the (fewer) free parameters, considerably reducing the computational time.

We observed that for the specific values of $\Delta$ considered, the procedure used is too rough (typically, we half the value of $\eta$ adaptively, which leads to a drastic slowing down for $\Delta =1.5$ at $L=24$); however, using other minimization algorithm as the Newton gradient descent, this obstacle is completely removed at $L=24$, and the results for $L\leq 10$ considered are slightly better. The $\eta$ control factor is replaced by the inverse of the matrix:
\begin{equation}
	\Xi_{\alpha,\beta}= \partial_\alpha \partial_\beta S(\rho|\sigma_\textup{BW}(\vec{w})).
\end{equation}
The main difference with respect to the previous algorithm is that the new parameters are proposed as:
\begin{equation}
	{w_\alpha}^{(n+1)} ={w_\alpha}^{(n)} + \sum_\beta \Xi_{\alpha,\beta}^{-1}\partial_\alpha S(\rho|\sigma_\textup{BW}(\vec{w}^{(n)})).
\end{equation}
Our point is that even a rough algorithm as the adaptive gradient descent presented here gives good quality results, up to very high accuracy (see Fig.~\ref{fig:figure333}).

It is worth to stress that the technique we propose is immediately adapted to cases where the EH ansatz is modified. As an example, we have carried out simulations utilizing the CFT ansatz~\cite{Cardy:2016aa} on the lattice~\cite{Giudici:2018aa}:
\begin{equation}
H_\textup{CFT}=\sum_{a,b=x,y,z}\sum_{r>0} \frac{L}{\pi}\sin\left(\pi \frac{r}{L}\right)\left(w_{a b}S_{a,r}S_{b,r+1}\right)
\end{equation}
with $w_{xx}=w_{yy}\neq w_{zz}\neq 0$ as the only non-zero parameters. The results are summarized in figure~\ref{fig:figure333}b. In practice, the reduced density matrix is described equally well by both the adapted CFT and BW result, and convergence to the correct solution is similar.

Finally, a comment is in order on the following problem: if the initial wave function is not exactly the ground state one (due, e.g., to imperfections in the measurement), but contains some admixture to low-lying states, how is the method coping with this? Answering this question requires the knowledge of the EH structure of low-lying excited states, which is available in (1+1)-d CFTs~\cite{Lashkari:2014aa}. In particular, the relative entropy between the GS and a low-lying excited state is finite, and proportional to the scaling dimension of the operator generating the excited state, for the specific half-partition we consider here. 

In the context of the XXZ model, we expect that, if fed not with the GS, but with a low-lying excited state, the algorithm will return a finite relative entropy (not necessarily with the same EH correspondent to the GS), indicating convergence has failed, everywhere apart from the vicinity of the ferromagnetic point. In the latter regime, the Luttinger parameter diverges, and since neutral excitations have dimension $\propto1/K$, the relative entropy between those excitations and the GS will be finite but small. As such, care must be taken when considering states with sizeable effects from excited states. Beyond (1+1)d CFTs, for a discussion on the relative entropy between ground and excited states in the context of holography, see Ref.~\cite{Blanco2013}.

\begin{figure}[h]
	\centering
	\includegraphics[width=0.60\columnwidth]{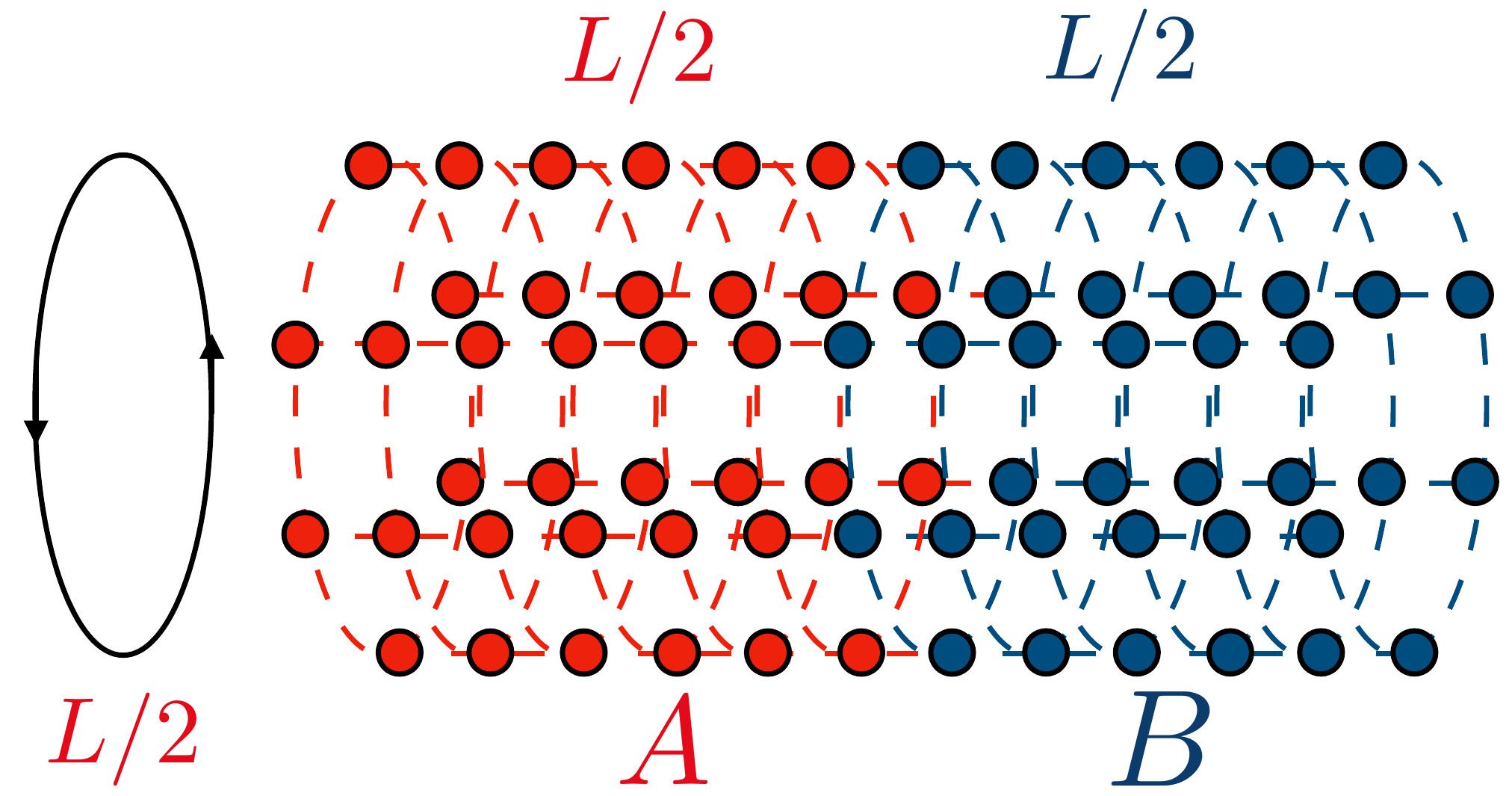}
	\includegraphics[width=0.35\columnwidth]{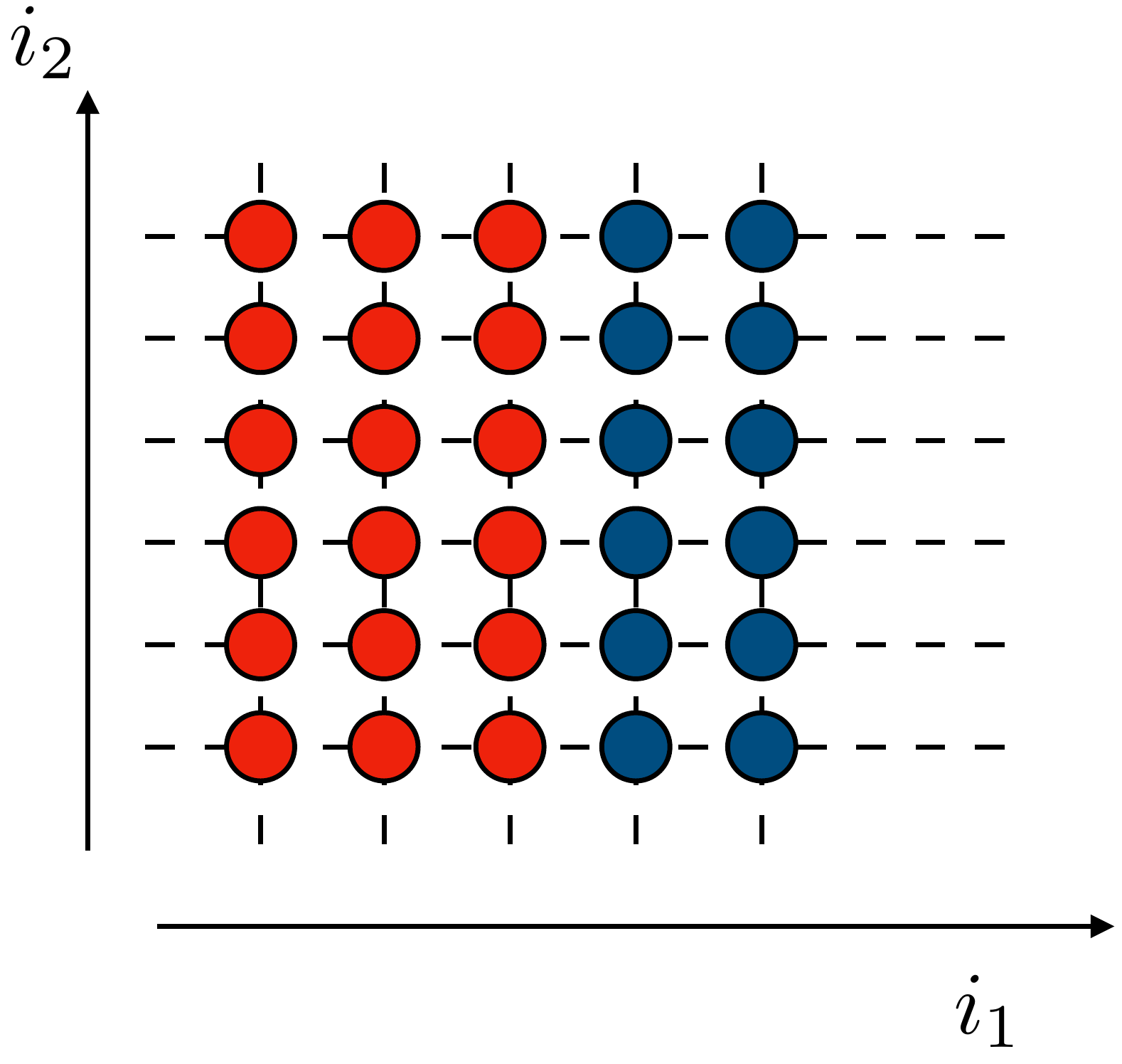}
	\caption{(Left) Partition used for the $2D$ model: $L/2\times L/2$. (Right) Direction $i_1$ is perpendicular to the cut (and start from the open boundary); $i_2$ is parallel to the cut and wraps around the cilinder.}
	\label{fig:figure2}
\end{figure}

\section{2D models: bilayer Heisenberg model}
\label{2d}
\begin{figure}[h]
	\centering
	\includegraphics[width=0.99\columnwidth]{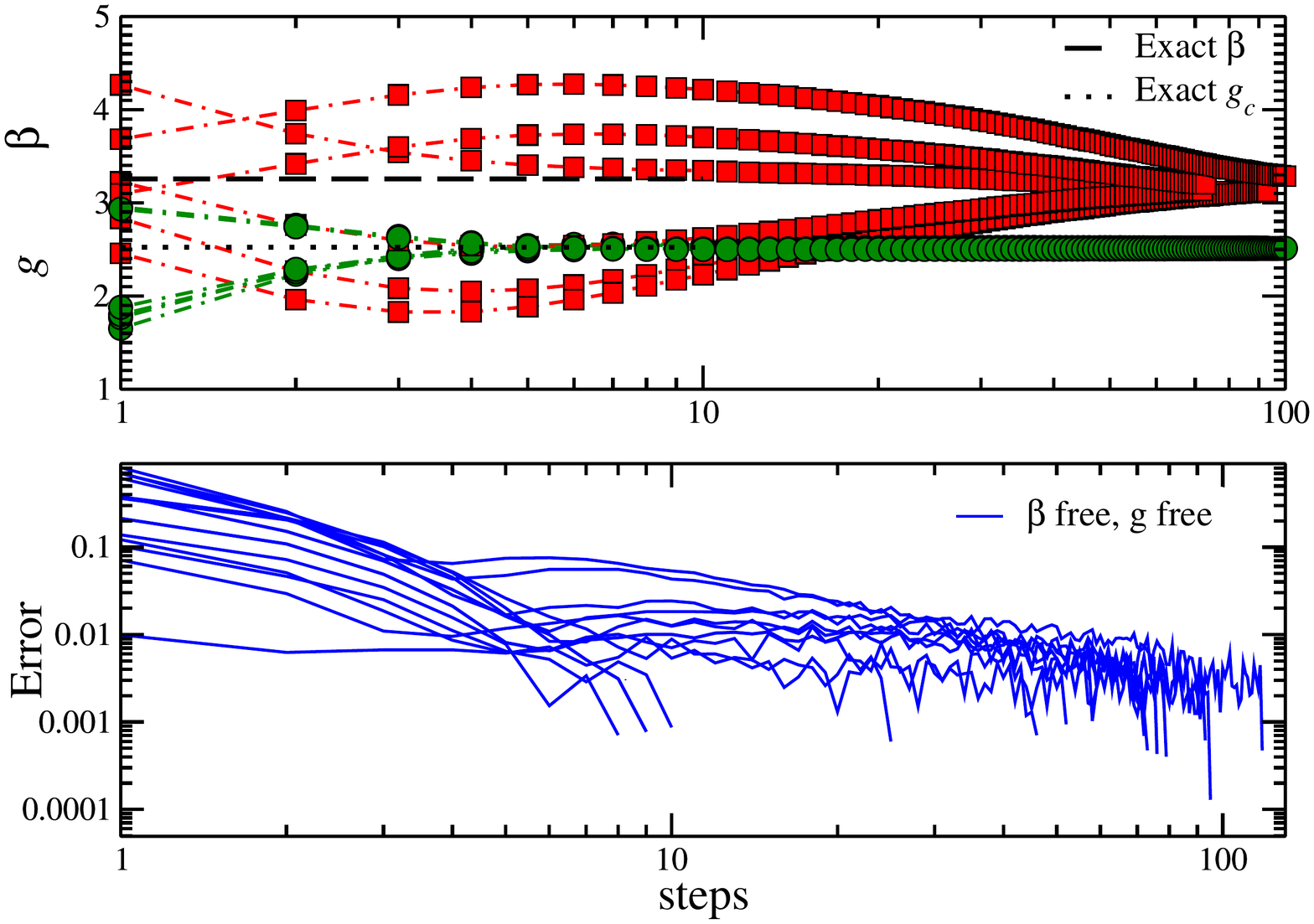}
	\caption{(Top) Convergence of the parameters at $L=8$. We see that the coupling constant converges faster that the entanglement temperature. (Bottom) Error trajectories as a function of steps.}
	\label{fig:figure3}
\end{figure}
Finally, we consider the bilayer Heisenberg model (HM)
\begin{eqnarray}\label{xxz2DH}
H_\textup{bilayer}&=&\sum_{l=1,2} \sum_{\langle \vec{i},\vec{j}\rangle} S_{\vec{i},l} S_{\vec{j},l} +g \sum_{\vec{i}}  S_{\vec{i},1} S_{\vec{i},2},
\end{eqnarray}
where $\vec{i}$ and $\vec{j}$ label the sites within the planes (square lattice), and $l$ are the label of the planes.
For $g=0$, the ground state of the  two uncoupled planes has antiferromagnetic (AFM) long-range order;
while  an AFM-Singlet quantum phase transition, whose  low-energy physics is described by a non-linear sigma model, takes place at $g=g_c$ \cite{sandvik1994}. 
We consider that the system has a cylinder geometry, with size $L\times L/2$;
Fig.~\ref{fig:figure2} illustrates the system and the partition considered here.  

As a check of the gradient descent procedure for 2D systems, we obtain in the main text the parent Hamiltonian of the ground state, $\ket{\Psi_0}$, of the bilayer HM at $g_c$.
To do so, first,  we use the quantum Monte Carlo (QMC) method Stochastic Series expansion  to obtain $\sum_r r \langle{O_{\alpha,r}}\rangle_{\Psi_0}$;
where $\langle{O_{\alpha,r}}\rangle$ is obtained for temperatures low enough to guarantee that the system already converged to the ground state (e.g., $\beta = 4L$).
Second, during the gradient descent minimization  part,
we consider the BW ansatz for the EH of the ground state half-bipartition 
\begin{eqnarray} \label{xxz2D}
H_\textup{BW}&=&\beta_c \sum_{l=1,2}\sum_{\vec{i},\delta} {i}_x S_{(i_x,i_y),l} S_{(i_x+\delta,i_y),l}  \nonumber\\  
&+& \beta_c \sum_{l=1,2} \sum_{\vec{i},\delta} \left({i}_x - 1/2 \right) S_{(i_x,i_y),l} S_{(i_x,i_y+\delta),l}  \nonumber\\
&+&\beta_c g\sum_{\vec{i}}  (i_x-1/2) S_{\vec{i},1} S_{\vec{i},2},
\end{eqnarray}
where $i_x > 0$, and $\beta_c$ is the ``entanglement inverse temperature''.
See Ref. \cite{Giudici:2018aa} for more details about the 2D version of the  BW-EH.
As we discuss in the main text (see Fig. \ref{fig:2DHeis}), we perform this minimization  by (i) fixing the value of $\beta_c$ to the exact one ($\beta_c = 3.307$) and considering the 
$g$ term as a free parameter, or by (ii)  considering both $\beta_c$ and $g$ as free parameters.
Fig. \ref{fig:figure3}, illustrates the convergence of $g$ and $\beta_c$ using the procedure (ii).

\end{document}